# FRAGMENTATION OF STABILITY DOMAINS OF DARK SOLITONS AND DARK BREATHERS AND DRIFTING SOLITONS AT HIGH PUMP INTENSITIES IN NORMAL DISPERSION KERR MICRORESONATORS

Valery E. Lobanov[1], Olga V. Borovkova[1,2], Alexander K. Vorobyev[1,3], Vladislav I. Pavlov[2,4], Dmitry A. Chermoshentsev[1,3,5], Igor A. Bilenko[1,2]

[1]*Russian Quantum Center, 143026, Skolkovo, Russia*
[2]*Faculty of Physics, Lomonosov Moscow State University, 119991, Moscow, Russia*
[3]*Moscow Institute of Physics and Technology, 141701, Dolgoprudny, Russia*
[4]*Russian Metrological Institute of Technical Physics and Radio Engineering, 141570, Moscow, Russia*
[5]*Skolkovo Institute of Science and Technology, Moscow 143025, Russia*

## Abstract

Stability domains (i.e. pump frequency detuning range) of a single dark soliton (or platicon) and dark breather in high-Q Kerr optical microresonators with normal group velocity dispersion is studied for a wide range of pump amplitudes within the framework of the Lugiato-Lefever model. The effect of the significant fragmentation of the stability domains at high pump intensities is revealed. The existence of stable drifting dark solitons (platicons) is demonstrated above the threshold pump amplitude value. Properties of drifting solitons are investigated.

## I. Introduction

The demonstration of generation of optical frequency combs in high-quality-factor optical microresonators [1-3] was a breakthrough discovery that significantly expanded the scope of frequency combs in various fields of science and technology. Even greater opportunities were opened up when the excitation of the coherent low-noise frequency combs in the form of dissipative Kerr solitons was discovered [4,5]. To date, the generation of such signals has been demonstrated in various microresonator platforms, bulk and integrated, and in various spectral ranges [3,5,6-13]. Dissipative solitons are successfully used in metrology [14], spectroscopy [15], astrophysics [16,17], telecommunication systems [18,19], quantum optics [20], lidars [21,22], microwave photonics [23] and many other areas [5,24]. It should be noted that the majority of results were obtained for bright solitons, which can be generated in microresonators at anomalous group velocity dispersion. For normal group velocity dispersion, there are dark solitons or platicons (flat-top solitonic pulses), which are the bound states of two switching waves propagating in opposite directions [25-31]. It was shown experimentally and numerically that the generation of dark solitons makes it possible to achieve a much higher efficiency of the pump energy conversion into the energy of the generated spectral components than the generation of bright solitons [28,32,33]. This can be used, for example, to design and build more efficient telecommunications systems [34,35]. Recent study also demonstrates the advantages of dark pulses over bright solitons in terms of quantum-limited coherence [36].

However, the generation of such signals is a much more difficult problem due to the almost complete absence of modulation instability in the case of normal group velocity dispersion. To solve this problem several methods were elaborated. First, one may use special microresonator structures, for example, multiresonator or photonic molecules, which make it possible to effectively control the dispersion [28,33,37-39]. Second, more complex types of pumping, two-frequency [40], amplitude-modulated [41,42] or pulsed [43] can be applied. Additionally, special regimes can be favorable for platicon generation, such as self-injection locking [44-47] or loss modulation [48] regimes. Furthermore, platicon excitation was shown to be possible due to the thermal effects [49] or multiphoton absorption and free-carrier effects [50].

Despite all the promise of using dark solitons, their properties have been studied much less than the properties of bright ones. One of the crucial issues for the practical application of solitons is their stability. For dark solitons, it was investigated in a number of works [26, 30, 31, 51] and both stable propagation regimes and oscillating (breathing) and chaotic regimes were shown. The influence of higher-order dispersion effects [52-58], Rayleigh [59] and Raman scattering [60-62] and thermal effects [63] was also studied. However, most of the results demonstrated were obtained for low pump amplitudes, which does not give a complete understanding of the applicability range of dark solitons. In our work, using Lugiato-Lefever model [26, 64, 65] we analyzed the stability domains of single-dip dark solitons and dark breathers [51] in terms of the detuning values of the pump frequency from the pumped mode eigenfrequency for a wide range of pump amplitudes. It was found that with an increase of this parameter, a significant spectral fragmentation of the stability range takes place, and at the pump strong enough several spectrally separated stability domains can be observed. In addition, it was found that stable drifting solitons exist at pump powers exceeding a certain value. Amplitude profiles of such solitons are asymmetric. The drift velocity, and hence the repetition rate of such solitons, is different for the solitons from different branches and depends both on the pump amplitude and on the pump frequency detuning. Stability of the reported drifting states was verified and confirmed by the linear stability analysis method.

## II. Model

For numerical analysis we used the Lugiato-Lefever equation [26] for the slowly varying envelope $\Psi$ of the intracavity field:

$$\frac{\partial \Psi}{\partial \tau} = i\frac{d_2}{2}\frac{\partial^2 \Psi}{\partial \varphi^2} - [1+i\alpha]\Psi + i|\Psi|^2\Psi + F. \qquad (1)$$

Here $\tau = \kappa t/2$ denotes the normalized time, $\kappa = \frac{\omega_0}{Q}$ is the cavity decay rate, $Q$ is the loaded quality factor, $\omega_0$ is the pumped mode resonant frequency, $\varphi \in [-\pi; \pi[$ is an azimuthal angle in a coordinate system rotating with the angular frequency equal to the microresonator free spectral range (FSR) $D_1$, $d_2 = 2D_2/\kappa$ is the group velocity dispersion (GVD) coefficient, positive for the anomalous GVD, negative for the normal GVD [the microresonator eigenfrequencies are assumed to be $\omega_\mu = \omega_0 + D_1\mu + \frac{1}{2}D_2\mu^2$, where $\mu$ is the mode number, calculated from the pumped mode]; $\alpha = 2(\omega_0 - \omega_p)/\kappa$ is the normalized detuning of the pump frequency $\omega_p$ from the pumped mode resonance. The normalized pump amplitude is $F = \sqrt{\dfrac{8\omega_p c n_2 \eta P_{in}}{\kappa^2 n^2 V_{eff}}}$, where $n$ s the refractive index of the pumped microresonator mode, $c$ is the speed of light in vacuum, $V_{eff}$ is the effective mode volume, $n_2$ is the nonlinear refractive index, $\eta$ is the coupling efficiency [$\eta = 1/2$ for critical coupling, $\eta \to 1$ for overloaded], $P_{in}$ is the input pump power.

First, we searched for the stationary solutions of Eq. (1) (when $\dfrac{\partial \Psi}{\partial \tau} = 0$) in the form of dark solitons or platicons by means of the relaxation method. Then the stability of obtained dark soliton solutions was tested by imposing random perturbations (additive and/or multiplicative) in the solutions and simulating their subsequent evolution up to $\tau = 10^4$ using Eq. (1). We integrated numerically the Eq. (1) making use of a split-step Fourier algorithm with periodic boundary conditions. Spatial grids from 1024 to 2048 points were used. It was checked that a further increase in the number of grid points does not change the results. The temporal step was lowered up to $h_\tau = 10^{-4}$ in order to yield $h_\tau$-independent results. Solitons that kept their shape during propagation were considered as the stable ones. Such analysis was performed for wide ranges of pump amplitude values up to $F = 12$. Note, that in Refs. [26, 30, 31] stability analysis was performed for $F \leq 5$. We set $d_2 = -0.02$ and checked that results were qualitatively the same for other values of $d_2$.

## III. Stability domains analysis

As it was shown in Refs. [30-31] several energy levels corresponding to stable dark solitons or platicons with different widths (and different number of oscillations in the dip) may exist in the same spectral range organized in a bifurcation structure known as collapsed snaking [66-70]. Upper energy levels correspond to wider platicons or smaller number of oscillations. Existence range of the dark solitons coincides with the bistability range (see the top left panel in Fig. 1). Upper dark soliton energy level starts from the high-detuning boundary of the nonlinear resonance curve (see upper black line and green curve at the left panel in Fig. 1, $U = \int_{-\pi}^{\pi} |\Psi|^2 d\varphi$). Dark solitons from this first energy level and solitons from the odd levels counted from it are predictably unstable. For small pump amplitudes (for $F < 2.5$ for considered parameters) dark solitons from even levels are stable. See the stability domains for $F = 2$ in the top left panel in Fig. 1, where stable energy levels are marked red, unstable are marked black and green line depicts nonlinear resonance curve. Further we will consider only these levels and the first stable level will be referred as upper one. It is clearly seen in the top left panel in Fig. 1 that stability range of the dark solitons is significantly narrower than their existence range.

With increase of the pump amplitude $F$ soliton existence domain (in terms of pump frequency detuning values) shifts to larger detuning values. Above some pump amplitude value at the upper level corresponding before to the stable stationary solitons the particular domains begin to appear (see blue line at the top right panel in Fig. 1 for $F = 2.5$). At these domains the patterns remain localized but experience oscillatory behavior and can be called dark soliton breathers [51]. In other stability domains the dark solitons remain stationary and stable. It was discussed in [30, 31] that the transition from the stationary dark soliton to the dark breather results from the supercritical Hopf bifurcation [71-72] that can be shown by careful analysis of the system's linear eigenvalues [30, 70].

While pump amplitude $F$ increases, breather domain at upper level becomes wider and then some breathers become unstable and instability domain appears at the upper level (see thin black line breaking thick blue one at the middle left panel in Fig. 1 for $F = 3.0$). Solitons from this instability domain transforms into homogeneous states upon propagation. In Refs. [30, 31, 70] it is shown that in this case breathers undergo a period-doubling route to chaos and the disappearance of the breather in favor of the homogeneous steady state may be related to the destruction of a chaotic attractor in a boundary crisis [73, 74]. The width of the instability range becomes larger with the increase of pump amplitude $F$ (compare middle left panel for $F = 3.0$ and middle right panel for $F = 4.0$ in Fig. 1).

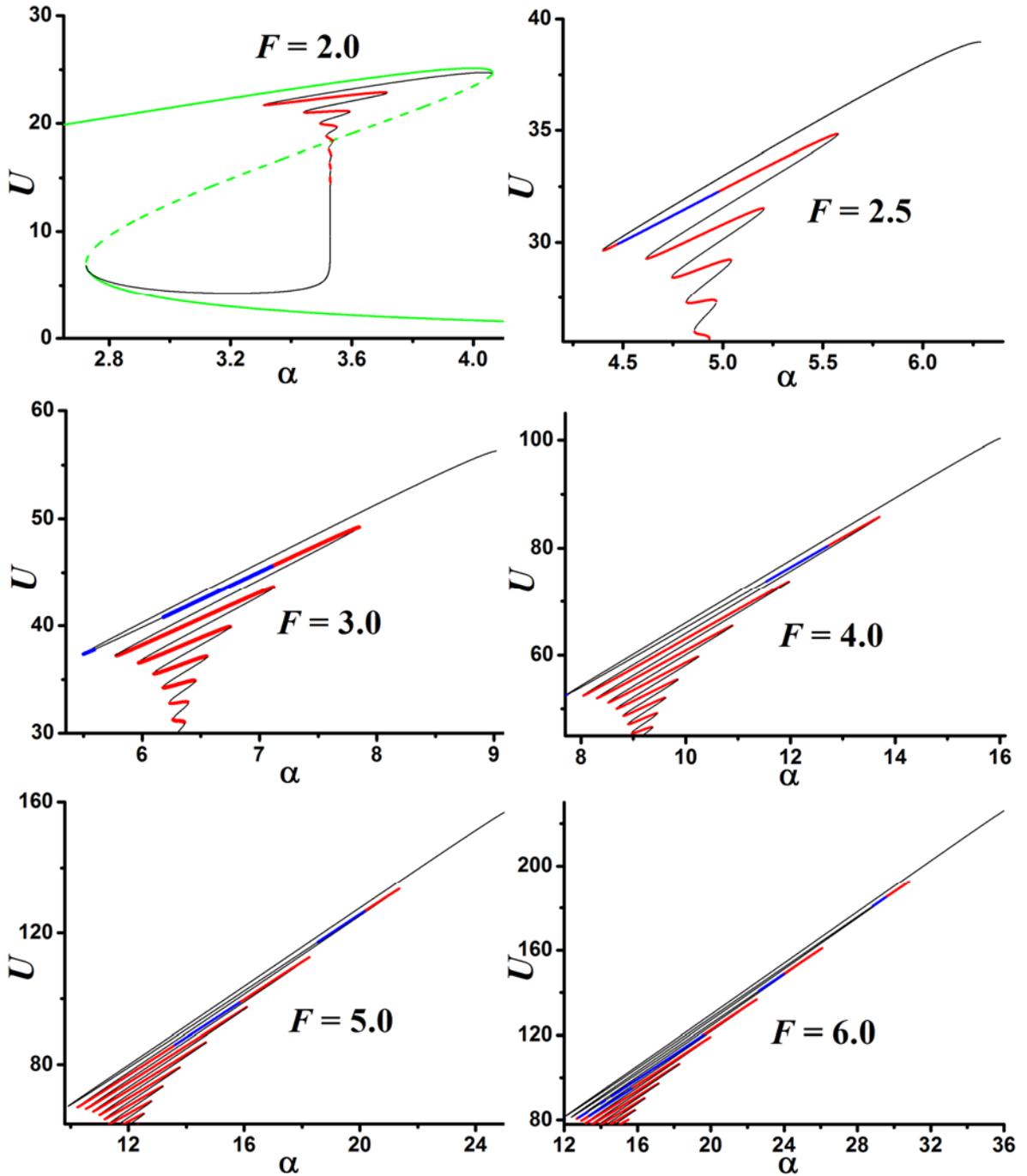

**Fig. 1.** The dependence of a single dark soliton energy on pump frequency detuning for $F = 2$ (top left panel), $F = 2.5$ (top right panel), $F = 3$ (middle left panel), $F = 4$ (middle right panel), $F = 5$ (bottom left panel) and $F = 6$ bottom (right panel). Thin black lines correspond to unstable solutions, thick red lines are used for stable dark solitons, thick blue lines stand for stable dark breathers. Green lines in the top left panel depicts the nonlinear resonance curve: solid lines for stable states, dashed for unstable. All quantities are plotted in dimensionless units.

With further increase of the pump amplitude $F$, breather and instability domains appear at lower previously stable levels (see bottom panels in Fig. 1). Number of such partially stable levels grows rapidly with the increase of the pump amplitude. For example, for $F = 5.0$ there are two such levels, for $F = 6.0$ we already have four levels, for $F = 7.0$ eleven such levels can be observed.

Interestingly, one can observed that even for high pump amplitudes at several upper levels there are adjoining breather and stable soliton domains at high-detuning boundary of each level (the first such domain can be seen at the middle right panel for $F = 4.0$ and bottom left panel for $F = 5.0$ in Fig. 1; at the bottom right panel in Fig. 1 for $F = 6.0$ there are two such domains). The spectral interval between these domains from different energy levels grows with the increase of pump amplitude that results in the pronounced fragmentation of the dark soliton stability domain at high pump amplitudes. It can be clearly seen in Fig. 2 where dark soliton stability map is given. The thick red lines depict detuning intervals for each pump amplitude value where stable solutions exist. Thin blue lines indicate the detuning ranges where only breathing solitons can be found.

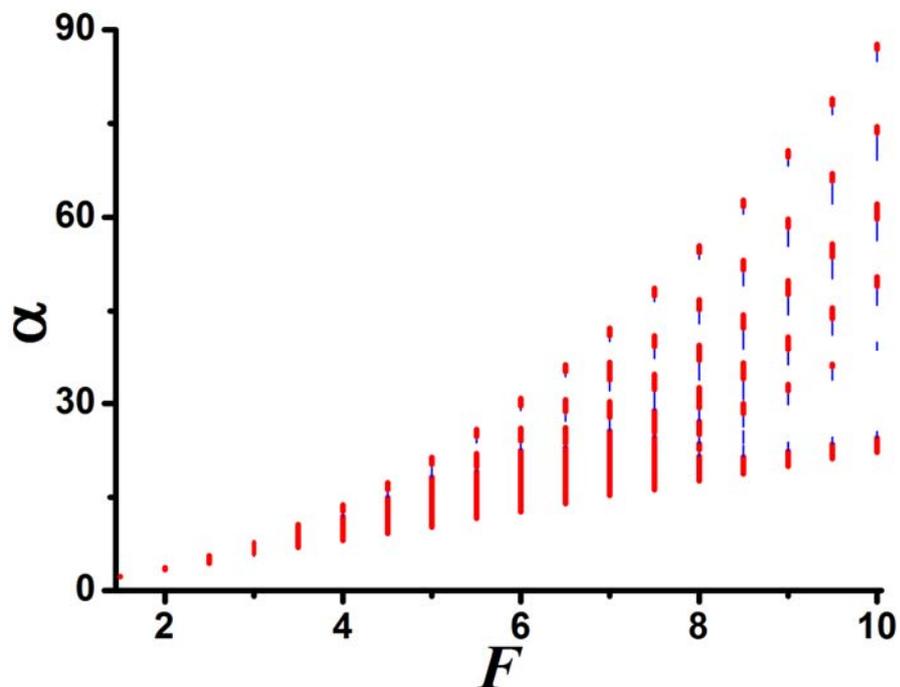

**Fig. 2.** Single dark soliton stability map. Thick red lines correspond to pump detuning interval when stable dark solitons can exist. Thin blue lines correspond to pump detuning ranges with dark breathers. All quantities are plotted in dimensionless units.

It should be noted that low-detuning stable soliton and breather domains shown in Fig. 2 for large $F$ consists of narrow domains from different energy levels. In Fig. 3 one can see stability domains for dark solitons and breathers from different energy levels for $F = 9.0$, the combination of which determines low-detuning domain shown in Fig. 2. The other five domains for $F = 9.0$ in Fig. 2 correspond to five upper levels.

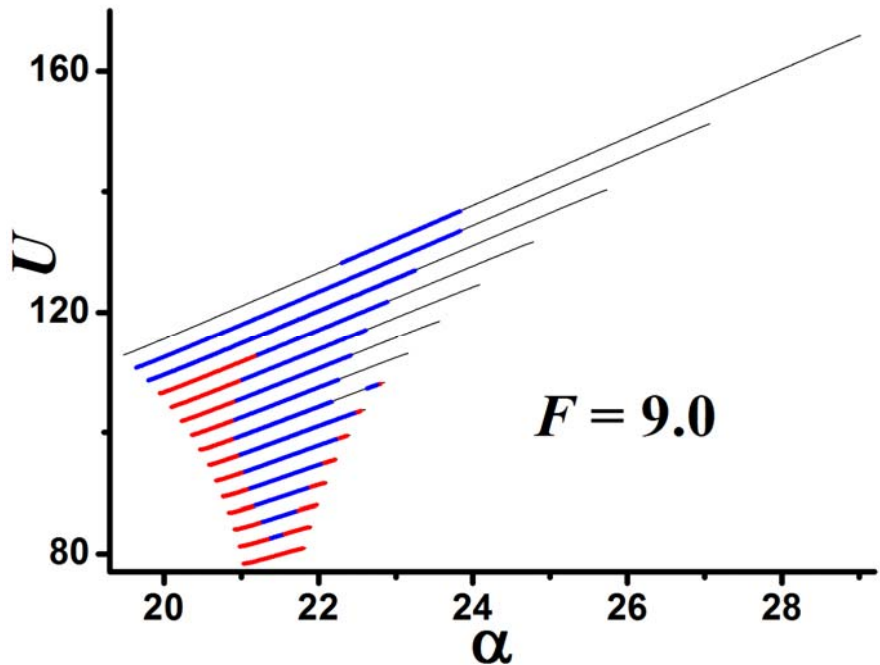

**Fig. 3.** The dependence of a single dark soliton energy on pump frequency detuning for lower energy levels defining low-detuning stability domain in Fig. 2 at $F = 9$. Thin black lines correspond to unstable solutions, thick red lines are used for stable solitons, thick blue lines stand for dark breathers. Unstable intermediate levels are omitted. All quantities are plotted in dimensionless units.

With further growth of pump amplitudes stability domains from low energy levels become more and more narrow and for $F = 12.0$ we found stable stationary dark solitons only at for upper levels (at $127.04 \leq \alpha \leq 127.72$; $109.0 \leq \alpha \leq 109.53$; $89.3 \leq \alpha \leq 91.86$; $73.75 \leq \alpha \leq 74.43$). Stable dark solitons and stable dark breathers in this case exist at $123.82 \leq \alpha \leq 127.72$; $102.3 \leq \alpha \leq 109.53$; $84.37 \leq \alpha \leq 91.86$; $70.61 \leq \alpha \leq 74.43$. It is clearly seen that separation between these domains is significant.

Note, that the obtained results were checked by the linear stability analysis approach. Within the framework of that method, the small perturbation to the soliton profile $\Psi_{sol}(\varphi)$ was considered in the form $\delta\Psi = u(\varphi)\exp(\lambda\tau) + v^*(\varphi)\exp(\lambda^*\tau)$ and using Eq. (1) a linearized system of equations was written:

$$\begin{cases} i\dfrac{d_2}{2}\dfrac{\partial^2 u}{\partial\varphi^2} - [1+i\alpha]u + 2i|\Psi_{sol}|^2 u + i\Psi_{sol}^2 v = \lambda u, \\ -i\dfrac{d_2}{2}\dfrac{\partial^2 v}{\partial\varphi^2} - [1-i\alpha]v - 2i|\Psi_{sol}|^2 v - i(\Psi_{sol}^*)^2 u = \lambda v. \end{cases} \quad (2)$$

Analysis of the eigenvalues $\lambda$ of the Jacobian of the linearized system (2) taking into account periodic boundary conditions confirmed the obtained results. In all cases for stable dark solitons the real part of $\lambda$ turned out to be nonpositive.

## IV. Drifting platicons

Surprisingly, it was noticed that in some cases at $F > 11.0$ unstable solitons may transform into stable drifting solitons upon propagation. To check that it is not a numerical artefact, we repeated the same simulations using different numerical grids (1024, 1400 and 2048 grid points) and obtained the same results. The results were also confirmed using coupled-mode equation system [75] solved in Matlab via variable step solver ode23 based on Bogacki–Shampine method where nonlinear terms were calculated using discrete Fourier transform [76]. Drifting dark solitons were also observed for different second-order dispersion coefficient values (at least from $d_2 = -0.01$ till $d_2 = -0.1$).

Similar drifting platicons at high pump amplitudes were also demonstrated numerically in the self-injection locking regime in presence of backscattering and nonlinear interaction with the backward wave [44]. Earlier platicon drift was also shown as a result of the third-order dispersion [53].

It should be noted that the effect of the soliton drift is well-known in different types of conservative and dissipative systems [77-86], including optical ones [87-

94]. It was shown that localized structures may lose their stability and start to move spontaneously as a result of the symmetry breaking drift bifurcation due to reasons, such as finite relaxation time [85, 88, 89] or delayed feedback [86, 92, 93] or Ising-Bloch transition [82, 83, 90, 91] and etc. Also, drift instability was shown for multi-frequency combs states in the form of spatiotemporal patterns in cylindrical Kerr microresonators with anomalous GVD [95]. Remarkably, recently nontrivial dynamics of two-dimensional frequency combs was reported in 2D microresonators at normal GVD [96].

Besides that, different symmetry breaking effects, including chiral symmetry breaking, has been already demonstrated in Kerr microresonators, and are actively studied nowadays [97-103]. In many cases such effects were shown to arise above a critical input power. Thus, the manifestation of symmetry-breaking-induced dark soliton dynamics is not unexpected at high pump intensities.

To study the origin and properties of drifting solutions in more detail the stationary form of Eq. (1) was rewritten using substitution $\theta = \varphi - V\tau,$ where $V$ is the normalized drift velocity:

$$i\frac{d_2}{2}\frac{\partial^2 \Psi}{\partial \theta^2} + V\frac{\partial \Psi}{\partial \theta} - [1+i\alpha]\Psi + i|\Psi|^2 \Psi + F = 0. \qquad (3)$$

Note, that the presence of the drift means the change of the dark soliton repetition rate equal to $\Delta f_{rep} = \frac{\kappa}{2}V$. We first verified that the drift structure profiles discussed above are solutions of Eq. (3). Then it was revealed that such drifting dark solitons (drifting platicons) can exist if pump amplitude exceeds some critical value. In our case they were observed for $F \geq 8.05$. Above this threshold it is possible to observe both stable stationary and stable drifting dark solitons (see Fig. 4).

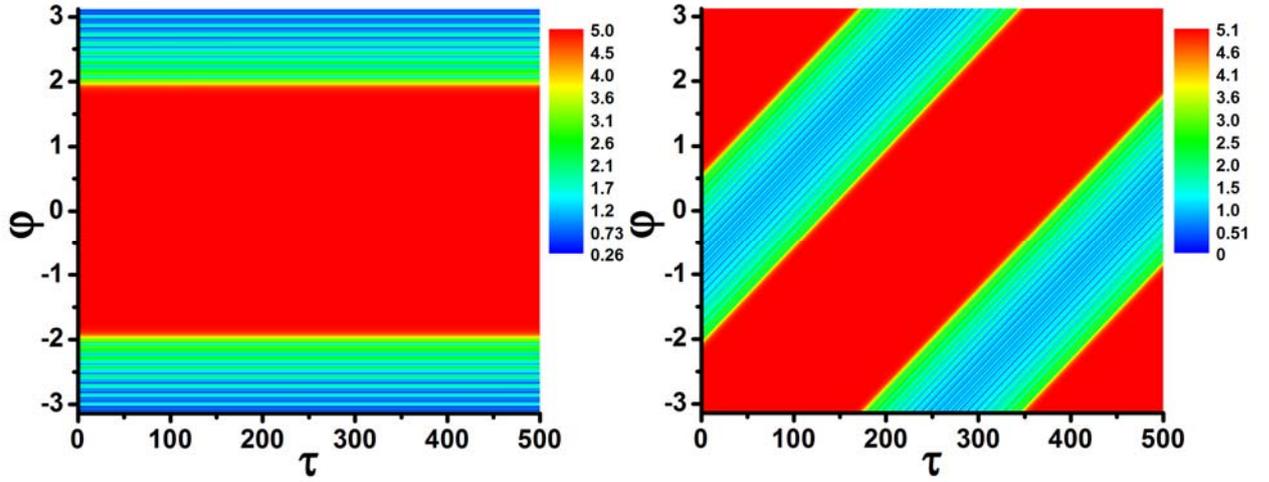

**Fig. 4.** Propagation dynamics of stable stationary platicon (dark soliton) at $F = 10$, $\alpha = 22.7$ (left panel) and stable drifting platicon at $F = 10$, $\alpha = 24$ (right panel). Color bars indicate field modulus. All quantities are plotted in dimensionless units.

However, for such high pump amplitudes the stability domains of the stationary platicons mostly becomes narrower with the growth of pump amplitude (except several upper levels discussed in the previous section), while the width of the detuning range providing the existence of the stable drifting dark solitons becomes larger with increase of $F$ (see Fig. 5).

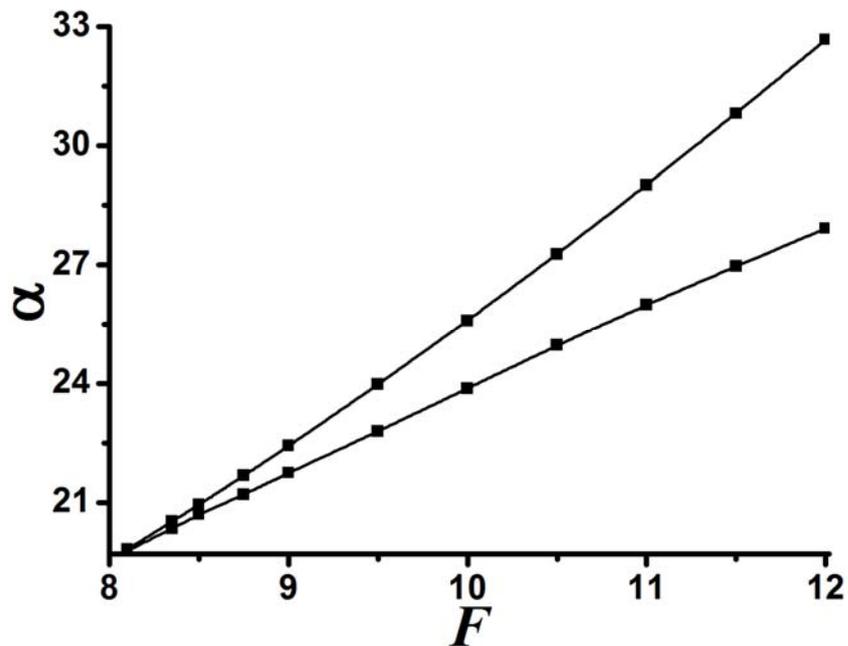

**Fig. 5.** Existence domain of stable drifting dark solitons (between black lines). All quantities are plotted in dimensionless units.

The structure of energy levels of drifting platicons is shown in the left panel in Fig. 6. At upper levels drifting breathers may also exists (thick segments). Drift velocity

$V$ depends on the detuning value $\alpha$ and decreases with the transition from upper levels to the lower ones (in the right panel in Fig. 6 one may see the dependences of the drift velocity on pump detuning for different energy levels; line colors in the right panel corresponds to the line color in the left panel).

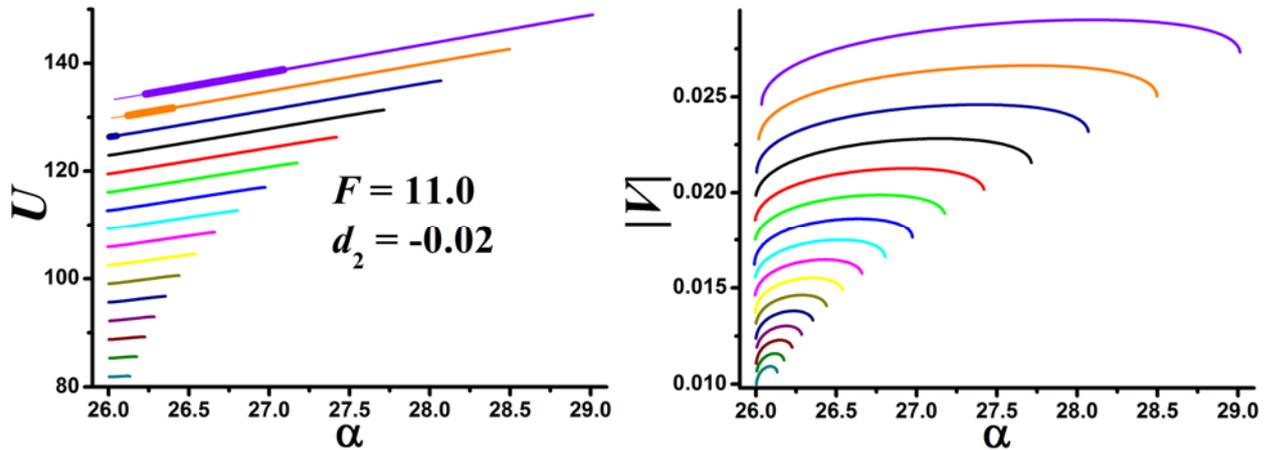

**Fig. 6.** The dependence of dark soliton energy (left panel) and drift velocity (right panel) on pump frequency detuning for stable drifting dark solitons at $F = 11$. Line colors in the right panel corresponds to the line colors in the left panel. Different line withs at upper levels in the left panel are used for different states: thin line for unstable states, medium for stable solitons, thick for dark breathers. All quantities are plotted in dimensionless units.

Amplitude profiles of the drifting platicons are slightly asymmetric (relatively the center of the dip, see Figs. 7 and 8). As it can be seen from Eq. (3), there may be a pair of mirror-symmetric drifting solitons with opposite velocities. Propagation dynamics is illustrated by Fig. 9 proving the stability of the drifting solitons.

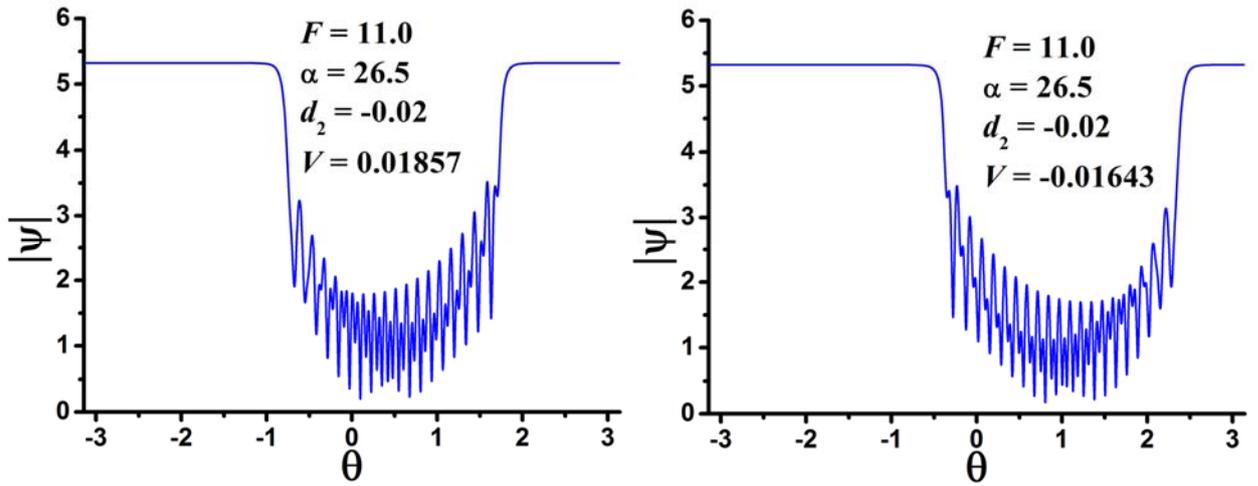

**Fig. 7.** Amplitude profiles of drifting platicons from different energy levels and with different drift velocity signs at $F=11$. Left panel corresponds to the seventh level from the top in Fig. 6, right one refers to the ninth level. Asymmetry of the solution with respect to the center of the dip is visible. All quantities are plotted in dimensionless units.

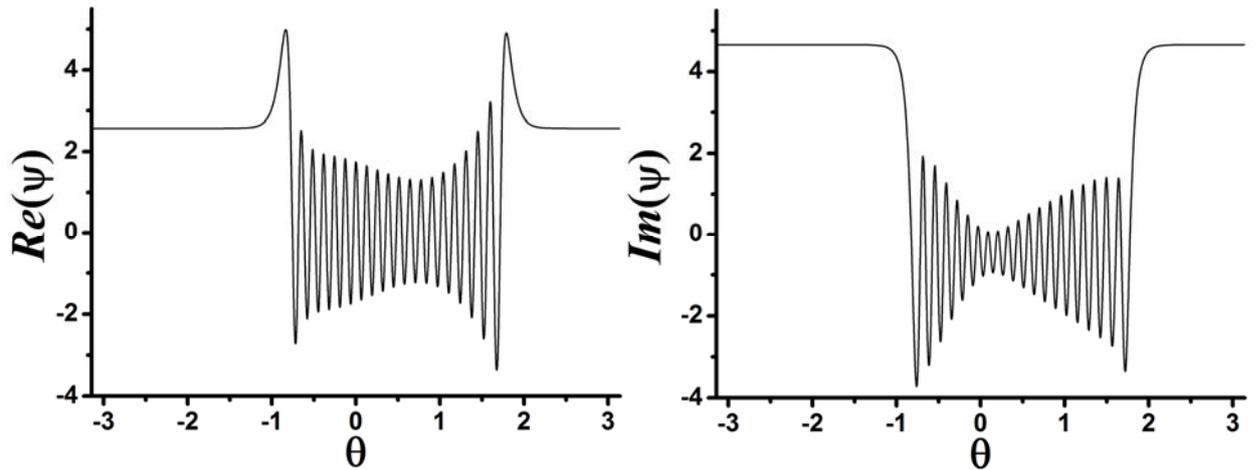

**Fig. 8.** Real and imaginary part profiles calculated at 2048-point grid for the stable drifting dark soliton shown in the left panel of Fig. 7 for $d_2 = -0.02$, $\alpha = 26.5$, $F = 11$. All quantities are plotted in dimensionless units.

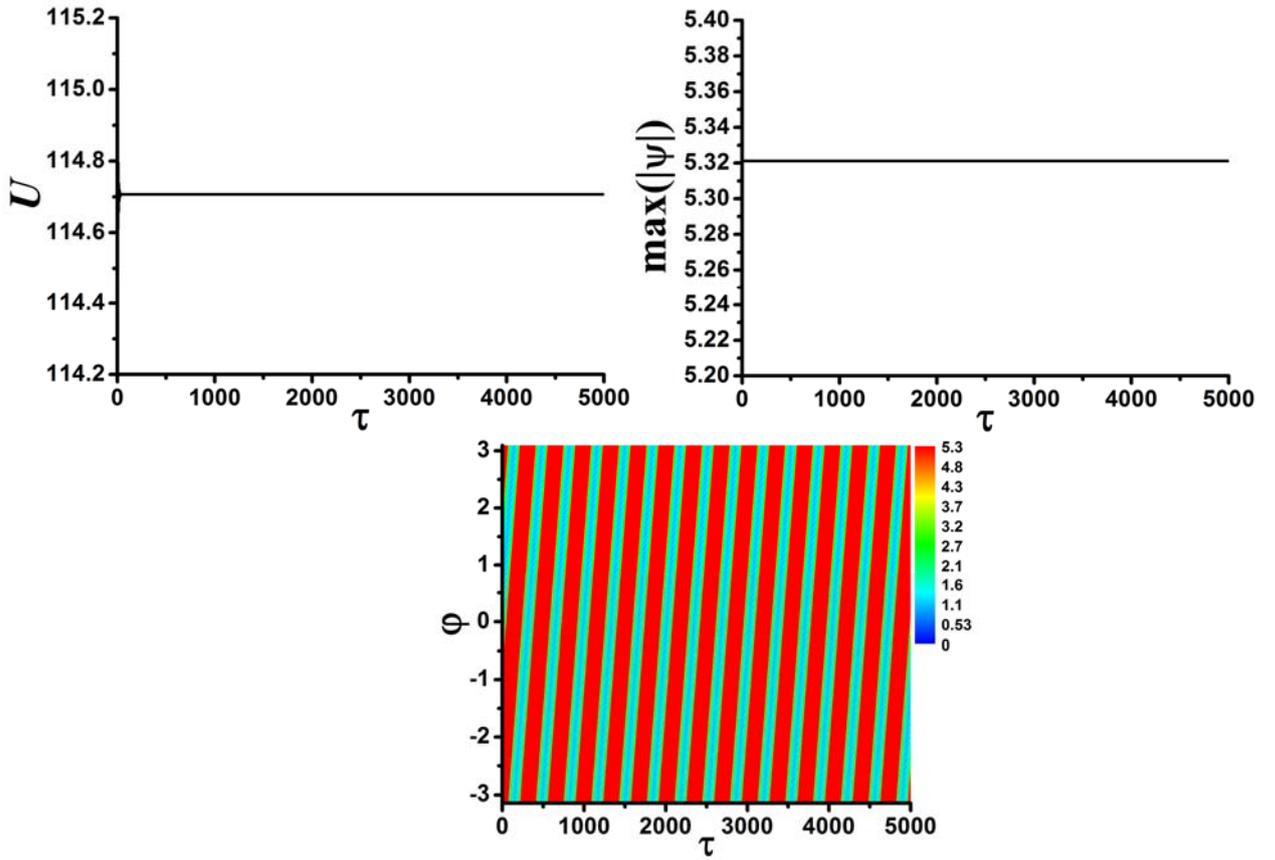

**Fig. 9.** Evolution dynamics of the drifting soliton shown in the left panel of Fig 7. (Left top panel) Soliton energy $U$ and (right top panel) soliton amplitude evolution upon propagation. (Bottom panel) Field modulus distribution evolution. Color bars indicate field modulus. All quantities are plotted in dimensionless units.

A careful examination of all possible solutions of Eq. (3) showed the tight connection between energy levels of drifting and resting solitons. First it was noticed that drifting solitons energy levels are situated between resting solitons levels. Then, it was found that the existence of drifting solution can be explained as a result of the specific symmetry-breaking bifurcation (see Fig. 10). It was revealed that above some critical pump amplitude value, in addition to intermediate monotonic branch (dotted black line in Fig. 10) connecting dark solitons levels (solid black lines in Fig. 10) and corresponding to unstable symmetric solitons, an additional S-shaped branch corresponding to asymmetric drifting solitons appears. Such branch can be divided into three segments: the lower one (brown line in Fig. 10) starting at the left end of the lower resting soliton level and the upper one (green line in Fig. 10) starting at the right end of the upper resting soliton level correspond to the unstable drifting solitons. The central segment of this branch (red line in Fig. 10) connects segments bifurcating from the resting solitons levels, and the corresponding drifting solitons can be stable.

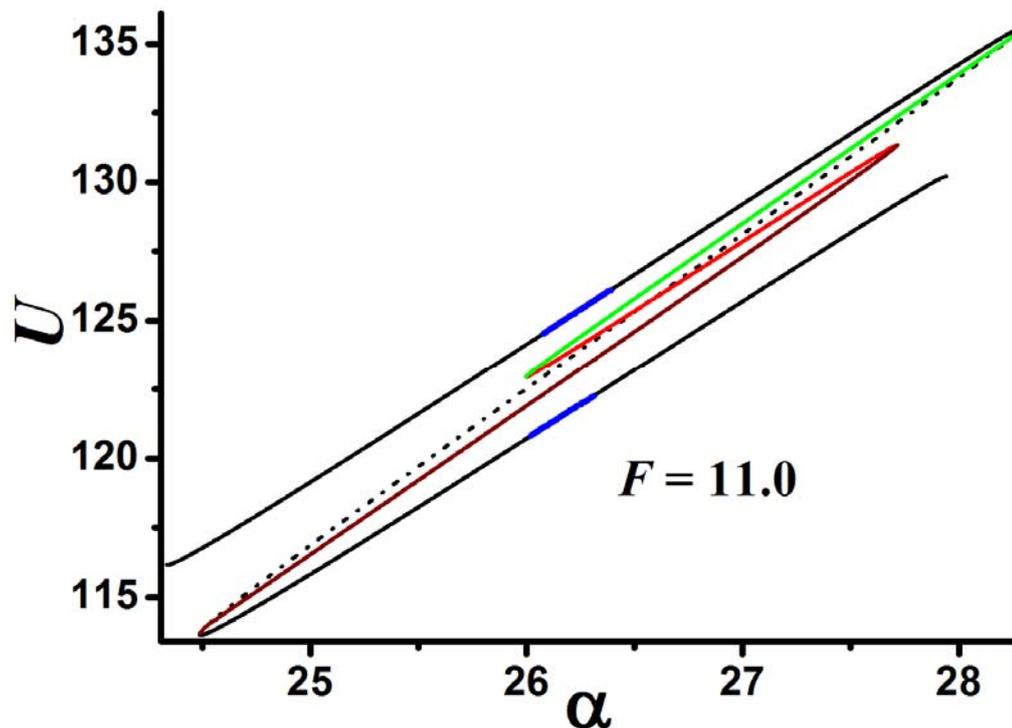

**Fig. 10.** Energy levels of resting and drifting platicons at $F = 11$. Solid black lines correspond to the energy levels of the potentially stable resting dark solitons, blue segments denote resting dark breathers stability domains, dotted black line depicts intermediate branch of unstable resting solitons. Green, brown, and red lines are segments of the S-shaped branch corresponding to the asymmetric drifting dark solitons. Solitons from the green and brown segments are unstable. All quantities are plotted in dimensionless units.

It is shown in Fig. 11 that drift velocity of the unstable platicons (brown and green lines) increases from zero when moving away from the bifurcation point. Drift velocity of stable platicons (red line) slightly depends on the pump frequency detuning.

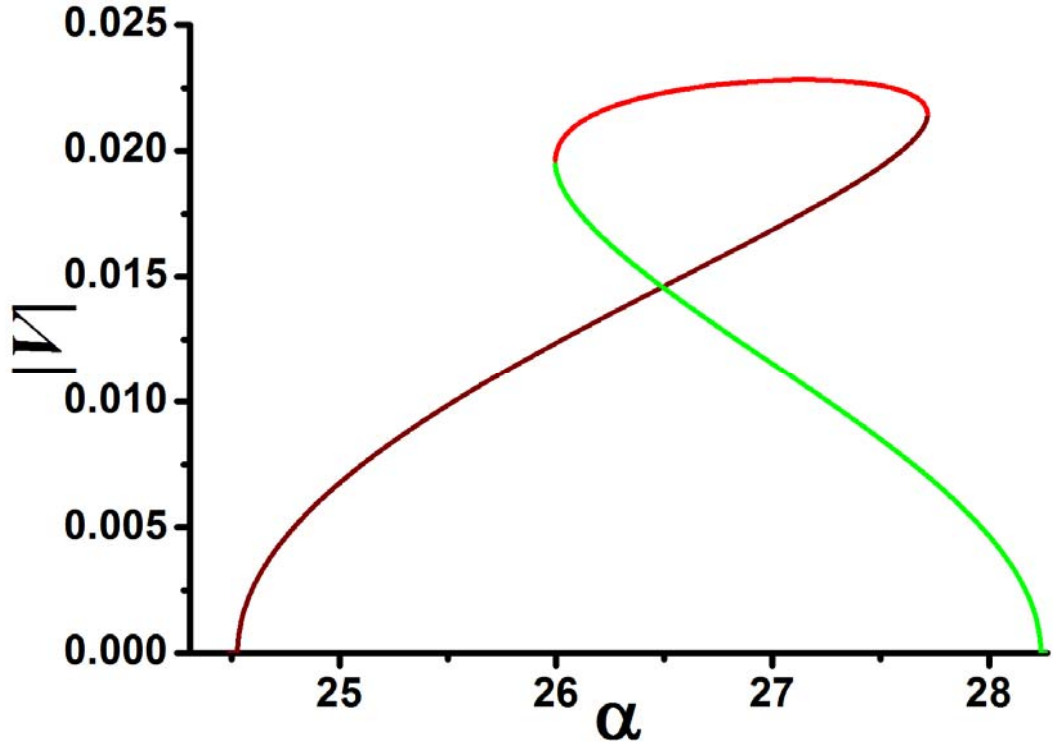

**Fig. 11.** The dependence of the drift velocity on pump frequency detuning at $F = 11$ for the drifting solitons corresponding to different segments of the S-shaped branch shown in Fig. 9. All quantities are plotted in dimensionless units.

Thus, it can be assumed, that the drifting states may emerge from a symmetry breaking pitchfork bifurcation taking place very close to the fold points of the collapsed snaking diagram. Their existence can be related to the asymmetric rung states known in the anomalous GVD regime (see Refs. [104] and [105]).

The existence of the stability range of the dark drifting solitons was confirmed via linear stability analysis method by means of the calculation of the eigenvalues of the Jacobian of the linearized system

$$\begin{cases} i\dfrac{d_2}{2}\dfrac{\partial^2 u}{\partial \theta^2} + V\dfrac{\partial u}{\partial \theta} - [1+i\alpha]u + 2i|\Psi_{sol}|^2 u + i\Psi_{sol}^2 v = \lambda u, \\ -i\dfrac{d_2}{2}\dfrac{\partial^2 v}{\partial \theta^2} + V\dfrac{\partial v}{\partial \theta} - [1-i\alpha]v - 2i|\Psi_{sol}|^2 v - i\left(\Psi_{sol}^*\right)^2 u = \lambda v, \end{cases} \quad (4)$$

that is a modification of system (2) used for the stability analysis of resting solitons. It was checked numerically that inside the stability domain for all eigenvalues $\operatorname{Re} \lambda \leq 0$.

It should be noted that stable drifting solitons without higher order effects are unknown and the reported ones in the anomalous GVD regime are unstable [105, 106].

It was also revealed that for the stability domains of the drifting solitons the same tendency takes place as for the resting dark solitons discussed in the previous section (see Fig. 1 for resting solitons and Fig. 12 for drifting solitons): with increase of the pump amplitudes the stationary solitons may transform into breathers that become unstable with further pump amplitude growth (compare stability domains for the same upper energy levels for different values of pump amplitude shown in Fig. 11). It is clearly seen that these processes start from the upper levels.

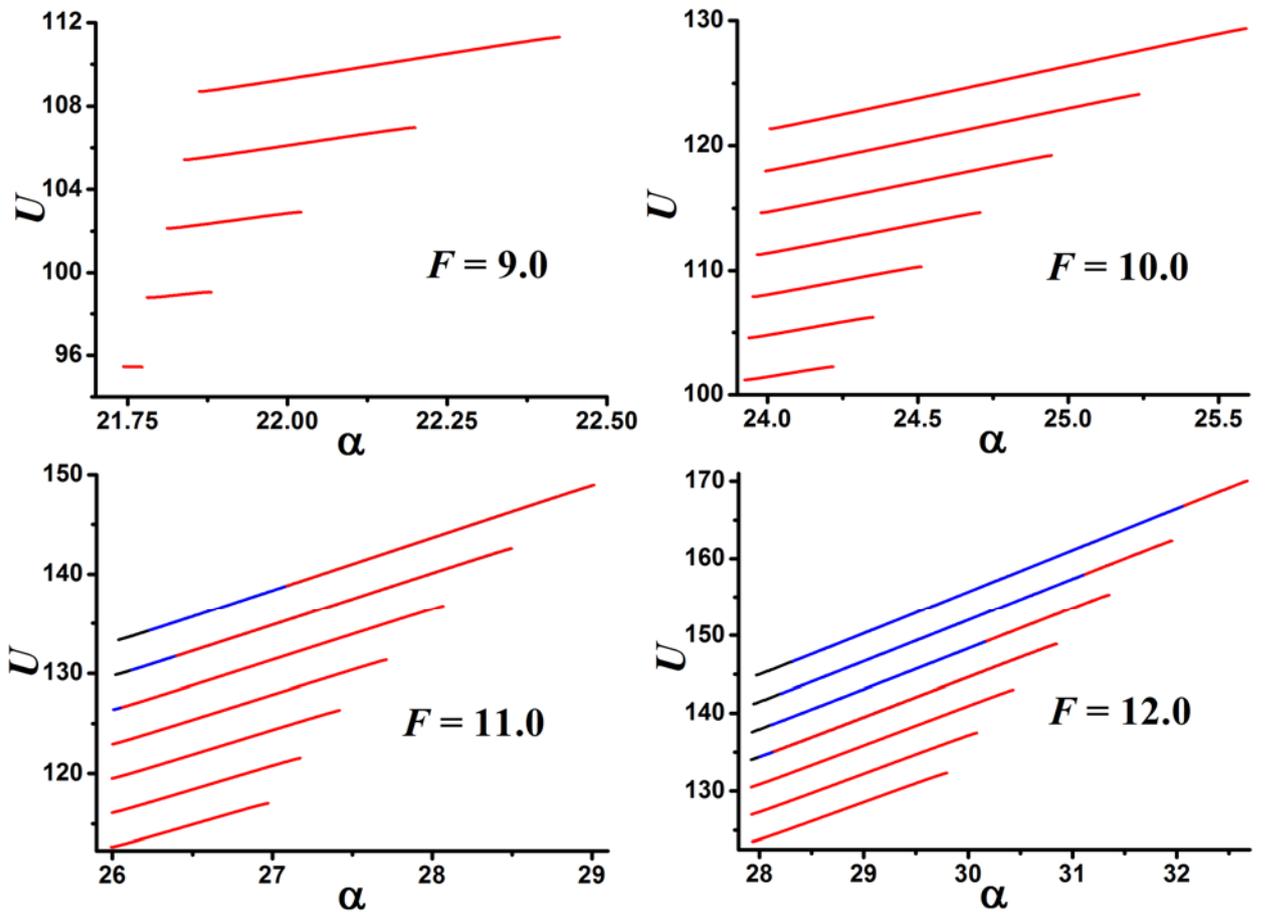

**Fig. 12.** The dependence of the drifting dark soliton energy on pump frequency detuning for $F = 9$ (top left panel), $F = 10$ (top right panel), $F = 11$ (bottom left panel), $F = 12$ (bottom right panel). Thin black lines correspond to unstable solutions, thick red lines are used for stable solitons, thick blue lines stand for dark breathers. All quantities are plotted in dimensionless units.

Drifting solitons were also studied for different GVD values $d_2$. It was revealed that their existence domain slightly depends on $d_2$ (small shift to the lower detuning values was observed). Drift velocity $V$ was found to increase with the growth of GVD coefficient absolute value (see Fig. 13).

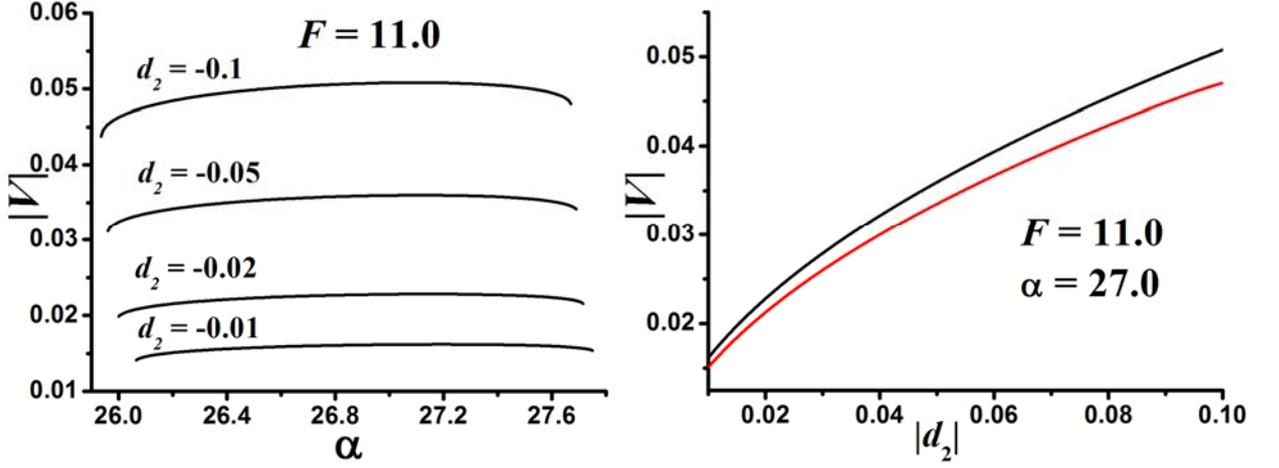

**Fig. 13.** (Left panel) The dependence of dark soliton drift velocity on pump frequency detuning for stable drifting dark solitons at $F = 11$ and different values of the GVD coefficient for the fourth energy level in Fig. 6. (Right panel) The dependence of dark soliton drift velocity on the absolute value of the GVD coefficient at $F = 11$, $\alpha = 27.0$ for the fourth (black) and fifth (red) energy levels in Fig. 6. All quantities are plotted in dimensionless units.

## V. Conclusion

The stability domains of dark solitons (platicons) and dark breathers in high-Q normal dispersion Kerr optical microresonators are studied for a wide range of pump amplitude within the framework of the Lugiato-Lefever model. The evolution of the stability regions with increasing pump amplitude is traced and the effect of the fragmentation of the stability domain at high pump amplitudes is revealed. The existence of stable drifting dark solitons (platicons) is demonstrated. It is found that stable drifting dark solitons exist if pump amplitude exceeds some threshold value. Stability of the drifting states is confirmed by the linear stability analysis approach. Note, that the existence of stable drifting solitons without higher-order effects in Kerr microresonator neither with anomalous GVD, nor with normal GVD has not been reported before. The connection between energy levels of drifting and resting solitons is revealed. It is shown that drifting states may emerge from a symmetry breaking pitchfork bifurcation taking place very close to the fold points of the collapsed snaking diagram. Amplitude profiles of the drifting platicons are slightly asymmetric (relatively the center of the dip) and there may be a pair of mirror-symmetric drifting solitons with opposite velocities. Drift velocity is found to increase with the growth of the GVD coefficient absolute value. The transformation of the stationary drifting solitons into dark breathers with increase of the pump amplitude is shown. Obtained results provide better understanding of the dark soliton applicability range in photonics and may be used as a starting point for further research taking into account high-order dispersion effects or backscattering (in the self-injection locking regime).

## VI. Acknowledgements

The work is supported by the Russian Science Foundation (project 23-42-00111). O.V.B. acknowledges the personal support from the Foundation for the Advancement of Theoretical Physics and Mathematics "BASIS."

## References

[1] P. Del'Haye, A. Schliesser, O. Arcizet, T. Wilken, R. Holzwarth, and T. J. Kippenberg, "Optical frequency comb generation from a monolithic microresonator," Nature **450(7173)**, 1214–1217 (2007).

[2] T. J. Kippenberg, R. Holzwarth, and S. A. Diddams, "Microresonator-based optical frequency combs," Science **332(6029)**, 555–559 (2011).

[3] A. Pasquazi, M. Peccianti, L. Razzari, D. J. Moss, S. Coen, M. Erkintalo, Y. K. Chembo, T. Hansson, S. Wabnitz, P. Del'Haye, X. Xue, A. M. Weiner, and R. Morandotti, "Micro-combs: A novel generation of optical sources," Phys. Rep. **729**, 1–81 (2018).

[4] T. Herr, V. Brasch, J. D. Jost, C. Y. Wang, N. M. Kondratiev, M. L. Gorodetsky, and T. J. Kippenberg, "Temporal solitons in optical microresonators," Nat. Photonics **8(2)**, 145–152 (2014).

[5] T. J. Kippenberg, A. L. Gaeta, M. Lipson, and M. L. Gorodetsky, "Dissipative Kerr solitons in optical microresonators," Science **361(6402)**, eaan8083 (2018).

[6] G. Lin, A. Coillet, and Y. K. Chembo, "Nonlinear photonics with high-Q whispering-gallery-mode resonators," Adv. Opt. Photonics **9(4)**, 828–890 (2017).

[7] A. L. Gaeta, M. Lipson, and T. J. Kippenberg, "Photonic-chip-based frequency combs," Nat. Photonics **13(3)**, 158–169 (2019).

[8] A. Kovach, D. Chen, J. He, H. Choi, A. H. Dogan, M. Ghasemkhani, H. Taheri, and A. M. Armani, "Emerging material systems for integrated optical Kerr frequency combs," Adv. Opt. Photonics **12(1)**, 135–222 (2020).

[9] M. Nie, Y. Xie, B. Li, and S.-W. Huang, "Photonic frequency microcombs based on dissipative Kerr and quadratic cavity solitons," Prog. Quantum Electron. **86**, 100437 (2022).

[10] C. Xiang, W. Jin, and J.E. Bowers, "Silicon nitride passive and active photonic integrated circuits: trends and prospects," Photon. Res. **10**, A82-A96 (2022).

[11] D. Zhu, L. Shao, M. Yu, R. Cheng, B. Desiatov, C. J. Xin, Y. Hu, J. Holzgrafe, S. Ghosh, A. Shams-Ansari, E. Puma, N. Sinclair, C. Reimer, M. Zhang, and M. Lončar, "Integrated photonics on thin-film lithium niobate," Adv. Opt. Photon. **13**, 242-352 (2021).

[12] X. Liu, A. W. Bruch, and H. X. Tang, "Aluminum nitride photonic integrated circuits: from piezo-optomechanics to nonlinear optics," Adv. Opt. Photon. **15**, 236-317 (2023).

[13] F. Baboux, G. Moody, and S. Ducci, "Nonlinear integrated quantum photonics with AlGaAs," Optica **10**, 917-931 (2023).

[14] S. B. Papp, K. Beha, P. Del'Haye, F. Quinlan, H. Lee, K. J. Vahala, and S. A. Diddams, "Microresonator frequency comb optical clock," Optica **1**, 10 (2014).

[15] M.-G. Suh, Q.-F. Yang, K. Y. Yang, X. Yi, and K. J. Vahala, "Microresonator soliton dual-comb spectroscopy," Science **354**, 600 (2016).

[16] M.-G. Suh, X. Yi, Y.-H. Lai, S. Leifer, I. S. Grudinin, G. Vasisht, E. C. Martin, M. P. Fitzgerald, G. Doppmann, J. Wang, D. Mawet, S. B. Papp, S. A. Diddams, C. Beichman, and K. Vahala, "Searching for exoplanets using a microresonator astrocomb," Nat. Photon. **13**, 25 (2019).

[17] E. Obrzud, M. Rainer, A. Harutyunyan, M. H. Anderson, M. Geiselmann, B. Chazelas, S. Kundermann, S. Lecomte, M. Cecconi, A. Ghedina, E. Molinari, F. Pepe, F. Bouchy F. Wildi, T. J. Kippenberg, and T. Herr, "A microphotonic astrocomb," Nat. Photon. **13**, 31 (2019).

[18] P. Marin-Palomo, J. N. Kemal, M. Karpov, A. Kordts, J. Pfeifle, M. H. P. Pfeiffer, P. Trocha, V. Brasch S. Wolf, M. H. Anderson, R. Rosenberger, K. Vijayan, W. Freude, T. J. Kippenberg, and C. Koos, "Microresonator-based solitons for massively parallel coherent optical communications," Nature **546**, 274 (2017).

[19] J. Feldmann, N. Youngblood, M. Karpov, H. Gehring, X. Li, M. Stappers, M. Le Gallo, X. Fu, A. Lukashchuk, A.S. Raja, J. Liu, C.D. Wright, A. Sebastian, T. J. Kippenberg, W. H. P. Pernice and H. Bhaskaran, "Parallel convolutional processing using an integrated photonic tensor core," Nature **589**, 52–58 (2021).

[20] M. Kues, C. Reimer, J. M. Lukens, W. J. Munro, A. M. Weiner, D. J. Moss, and R. Morandotti, "Quantum optical microcombs," Nat. Photonics **13**, 170 (2019).

[21] M.-G. Suh, and K. J. Vahala, "Soliton microcomb range measurement," Science **359**, 884–887 (2018).

[22] J. Riemensberger, A. Lukashchuk, M. Karpov, W. Weng, E. Lucas, J. Liu, and T. J. Kippenberg, "Massively parallel coherent laser ranging using a soliton microcomb," Nature (London) **581**, 164 (2020).

[23] J. Liu, E. Lucas, A. S. Raja, J. He, J. Riemensberger, R. Ning Wang, M. Karpov, H. Guo, R. Bouchand, T. J. Kippenberg "Photonic microwave generation in the X- and K-band using integrated soliton microcombs," Nat. Photonics **14**, 486–491 (2020).

[24] Y. Sun, J. Wu, M. Tan, X. Xu, Y. Li, R. Morandotti, A. Mitchell, and D. J. Moss, "Applications of optical microcombs," Adv. Opt. Photonics **15**, 86 (2023).

[25] W. Liang, A. A. Savchenkov, V. S. Ilchenko, D. Eliyahu, D. Seidel, A. B. Matsko, and L. Maleki, "Generation of a coherent near-infrared Kerr frequency comb in a monolithic microresonator with normal GVD," Opt. Lett. **39**, 2920 (2014).

[26] C. Godey, I. V. Balakireva, A. Coillet, and Y. K. Chembo, "Stability analysis of the spatiotemporal Lugiato-Lefever model for Kerr optical frequency combs in the anomalous and normal dispersion regimes," Phys. Rev. A **89**, 063814 (2014).

[27] X. Xue, Y. Xuan, Y. Liu, P.-H. Wang, S. Chen, J. Wang, D. E. Leaird, M. Qi, and A. M.Weiner, "Mode-locked dark pulse Kerr combs in normal-dispersion microresonators," Nat. Photonics **9**, 594 (2015).

[28] V. E. Lobanov, G. Lihachev, T. J. Kippenberg, and M. L. Gorodetsky, "Frequency combs and platicons in optical microresonators with normal GVD," Opt. Express **23**, 7713 (2015).

[29] X. Xue, M. Qi, and A. M.Weiner, "Normal-dispersion microresonator Kerr frequency combs," Nanophotonics **5**, 244 (2016).

[30] P. Parra-Rivas, E. Knobloch, D. Gomila, and L. Gelens, "Dark solitons in the Lugiato-Lefever equation with normal dispersion," Phys. Rev. A **93**, 063839 (2016).

[31] P. Parra-Rivas, D. Gomila, E. Knobloch, S. Coen, and L. Gelens, "Origin and stability of dark pulse Kerr combs in normal dispersion resonators," Opt. Lett. **41**, 2402 (2016).

[32] X. Xue, P.-H. Wang, Y. Xuan, M. Qi, and A. M. Weiner, "Microresonator Kerr frequency combs with high conversion efficiency," Laser Photon. Rev. **11**, 1600276 (2017).

[33] B. Y. Kim, Y. Okawachi, J. K. Jang, M. Yu, X. Ji, Y. Zhao, C. Joshi, M. Lipson, and A. L. Gaeta, "Turn-key, high-efficiency Kerr comb source," Opt. Lett. **44**, 4475 (2019).

[34] A. Fülöp, M. Mazur, A. Lorences-Riesgo, Ó. B. Helgason, P.-H. Wang, Y. Xuan, D. E. Leaird, M. Qi, P. A. Andrekson, A. M. Weiner, and V. Torres-Company, "High-order coherent communications using mode-locked dark-pulse Kerr combs from microresonators," Nat. Commun. **9**, 1598 (2018).

[35] Ó. B. Helgason, A. Fülöp, J. Schröder, P. A. Andrekson, A. M. Weiner, and V. Torres-Company, Superchannel engineering of microcombs for optical communications, J. Opt. Soc. Am. B **36**, 2013 (2019).

[36] C. Lao, X. Jin, L. Chang, H. Wang, Z. Lv, W. Xie, H. Shu, X. Wang, J. E. Bowers and Q.-F. Yang, "Quantum decoherence of dark pulses in optical microresonators," Nat Commun. **14**, 1802 (2023).

[37] X. Xue, Y. Xuan, P.-H. Wang, Y. Liu, D. E. Leaird, M. Qi, and A. M. Weiner, "Normal-dispersion microcombs enabled by controllable mode interactions," Laser Photonics Rev. **9**, L23 (2015).

[38] Ó. B. Helgason, F. R. Arteaga-Sierra, Z. Ye, K. Twayana, P. A. Andrekson, M. Karlsson, J. Schröder, V. Torres-Company, "Dissipative solitons in photonic molecules," Nat. Photonics **15**, 305–310 (2021).

[39] I. Rebolledo-Salgado, C. Quevedo-Galán, Ó.B. Helgason, A. Lööf, Z. Ye, F. Lei, J. Schröder, M. Zelan, V. Torres-Company, "Platicon dynamics in photonic molecules," Commun. Phys. **6**, 303 (2023).

[40] V. E. Lobanov, G. Lihachev, and M. L. Gorodetsky, "Generation of platicons and frequency combs in optical microresonators with normal GVD by modulated pump," EPL **112** 54008 (2015).

[41] V. E. Lobanov, N. M. Kondratiev, A. E. Shitikov, R. R. Galiev, and I. A. Bilenko, "Generation and dynamics of solitonic pulses due to pump amplitude modulation at normal group-velocity dispersion," Phys. Rev. A **100**, 013807 (2019).

[42] H. Liu, S.-W. Huang, W. Wang, J. Yang, M. Yu, D.-L. Kwong, P. Colman, and C. W. Wong, "Stimulated generation of deterministic platicon frequency microcombs," Photonics Res. **10**, 1877 (2022).

[43] Y. Xu, A. Sharples, J. Fatome, S. Coen, M. Erkintalo, and S. G. Murdoch, "Frequency comb generation in a pulse-pumped normal dispersion Kerr mini-resonator," Opt. Lett. **46**, 512-515 (2021).

[44] N. M. Kondratiev, V. E. Lobanov, E. A. Lonshakov, N. Y. Dmitriev, A. S. Voloshin, and I. A. Bilenko, "Numerical study of solitonic pulse generation in the self-injection locking regime at normal and anomalous group velocity dispersion," Opt. Express **28**, 38892 (2020).

[45] W. Jin, Q.-F. Yang, L. Chang, B. Shen, H. Wang, M. A. Leal, L. Wu, M. Gao, A. Feshali, M. Paniccia, K. J. Vahala, and J. E. Bowers, "Hertz-linewidth

semiconductor lasers using CMOS-ready ultra-high-Q microresonators," Nat. Photon. **15**, 346 (2021).

[46] G. Lihachev, W. Weng, J. Liu, L. Chang, J. Guo, J. He, R. Ning Wang, M. H. Anderson, Y. Liu, J. E. Bowers and T. J. Kippenberg, "Platicon microcomb generation using laser self-injection locking," Nat Commun **13**, 1771 (2022).

[47] N. M. Kondratiev, V. E. Lobanov, A. E. Shitikov, R. R. Galiev, D. A. Chermoshentsev, N. Yu. Dmitriev, A. N. Danilin, E. A. Lonshakov, K. N. Min'kov, D. M. Sokol, S. J. Cordette, Y.-H. Luo, W. Liang, J. Liu and I. A. Bilenko, "Recent advances in laser self-injection locking to high-Q microresonators," Front. Phys. **18**, 21305 (2023).

[48] M. Liu, Y. Dang, H. Huang, Z. Lu, Y. Wang, Y. Cai, and W. Zhao, "Loss modulation assisted solitonic pulse excitation in Kerr resonators with normal group velocity dispersion," Opt. Express **30**, 30176-30186 (2022).

[49] V.E. Lobanov, N.M. Kondratyev, and I.A. Bilenko, "Thermally Induced Generation of Platicons in Optical Microresonators," Optics Letters, **46(10)**, 2380-2383 (2021).

[50] M. Liu, L. Wang, Q. Sun, S. Li, Z. Ge, Z. Lu, W. Wang, G. Wang, W. Zhang, X. Hu, and W. Zhao, "Influences of multiphoton absorption and free-carrier effects on frequency-comb generation in normal dispersion silicon microresonators," Photonics Res. **6(4)**, 238–243 (2018).

[51] C. Bao, Y. Xuan, C. Wang, A. Fülöp, D. E. Leaird, V. Torres-Company, M. Qi, and A. M. Weiner, "Observation of Breathing Dark Pulses in Normal Dispersion Optical Microresonators," Phys. Rev. Lett. **121**, 257401 (2018).

[52] Y. He, S. Wang, X. Zeng, "Dynamics of Dispersive Wave Emission From Dark Solitons in Kerr Frequency Combs," IEEE Photon. J. **8**, 7102508 (2016).

[53] V. E. Lobanov, A. V. Cherenkov, A. E. Shitikov, I. A. Bilenko and M. L. Gorodetsky, "Dynamics of platicons due to third-order dispersion," Eur. Phys. J. D, **71(7)**, 185 (2017).

[54] P. Parra-Rivas, D. Gomila, L. Gelens, "Coexistence of stable dark- and bright-soliton Kerr combs in normal-dispersion resonators," Phys. Rev. A **95**, 053863 (2017).

[55] J. H. T. Mbé and Y. K. Chembo, "Coexistence of bright and dark cavity solitons in microresonators with zero, normal, and anomalous group-velocity dispersion: a switching wave approach," J. Opt. Soc. Am. B **37**, A69-A74 (2020).

[56] P. Parra-Rivas, S. Hetzel, Y. V. Kartashov, P. Fernández de Córdoba, J. Alberto Conejero, A. Aceves, and C. Milián, "Quartic Kerr cavity combs: bright and dark solitons," Opt. Lett. **47**, 2438-2441 (2022).

[57] M. H. Anderson, W. Weng, G. Lihachev, A. Tikan, J. Liu, and T. J. Kippenberg, "Zero dispersion Kerr solitons in optical microresonators," Nat. Commun. **13**, 4764 (2022).

[58] E. K. Akakpo, M. Haelterman, F. Leo, and P. Parra-Rivas, "Emergence of collapsed snaking related dark and bright Kerr dissipative solitons with quartic-quadratic dispersion," Phys. Rev. E **108**, 014203 (2023).

[59] V. E. Lobanov, A. E. Shitikov, R. R. Galiev, K. N. Min'kov, and N. M. Kondratiev, "Generation and properties of dissipative Kerr solitons and platicons in optical microresonators with backscattering," Opt. Express 28, 36544-36558 (2020).

[60] A. V. Cherenkov, N. M. Kondratiev, V. E. Lobanov, A. E. Shitikov, D. V. Skryabin, and M. L. Gorodetsky, "Raman-Kerr frequency combs in microresonators with normal dispersion," Opt. Express **25**, 31148-31158 (2017).

[61] P. Parra-Rivas, S. Coulibaly, M. G. Clerc, and M. Tlidi, "Influence of stimulated Raman scattering on Kerr domain walls and localized structures," Phys. Rev. A **103**, 013507 (2021).

[62] M. Liu, H. Huang, Z. Lu, Y. Wang, Y. Cai, and W. Zhao, "Dynamics of dark breathers and Raman-Kerr frequency combs influenced by high-order dispersion," Opt. Express **29**, 18095-18107 (2021)

[63] V. E. Lobanov, N. M. Kondratiev, A. E. Shitikov, O. V. Borovkova, S. J. Cordette, and I. A. Bilenko, "Platicon stability in hot cavities," Opt. Lett. **48**, 2353-2356 (2023).

[64] L. A. Lugiato and R. Lefever, "Spatial Dissipative Structures in Passive Optical Systems," Phys. Rev. Lett. **58**, 2209 (1987).

[65] Y. K. Chembo and C. R. Menyuk, "Spatiotemporal Lugiato-Lefever formalism for Kerr-comb generation in whispering gallery-mode resonators," Phys. Rev. A **87**, 053852 (2013).

[66] J. Knobloch, T. Wagenknecht, "Homoclinic snaking near a heteroclinic cycle in reversible systems," Physica D: Nonlinear Phenomena, **206(1)**, 82–93 (2005).

[67] A. Yochelis, J. Burke, E. Knobloch, "Reciprocal oscillons and nonmonotonic fronts in forced nonequilibrium systems," Phys. Rev. Lett., **97(25)**, 254501 (2006).

[68] J. Burke, A. Yochelis, and E. Knobloch, "Classification of spatially localized oscillations in periodically forced dissipative systems," SIAM J. Appl. Dyn. Syst. **7**, 651 (2008).

[69] Y.-P. Ma, J. Burke, and E. Knobloch, "Defect-mediated snaking: A new growth mechanism for localized structures," Phys. D (Amsterdam) **239**, 1867 (2010).

[70] P. Parra-Rivas, E. Knobloch, L. Gelens, D. Gomila, "Origin, bifurcation structure and stability of localized states in Kerr dispersive optical cavities," IMA Journ. Appl. Math**. 86(5)**, 856–895 (2021).

[71] B. D. Hassard, N. D. Kazarinoff, Y.-H. Wan. *Theory and Applications of Hopf Bifurcation* (New York: Cambridge University Press, 1981).

[72] S. Strogatz. *Nonlinear Dynamics and Chaos* (Reading, MA: Addison-Wesley, 1994).

[73] C. Grebogi, E. Ott, J. A. Yorke, "Crises, sudden changes in chaotic attractors, and transient chaos," Physica D: Nonlinear Phenomena, **7(1)**, 181–200 (1983).

[74] R. Hilborn, *Chaos and Nonlinear Dynamics: An introduction for Scientists and Engineers* (Oxford University Press, Oxford, 2000).

[75] Y. K. Chembo and N. Yu, "Modal expansion approach to optical-frequency-comb generation with monolithic whispering-gallery-mode resonators, Phys. Rev. A **82**, 033801 (2010).

[76] T. Hansson, D. Modotto, and S. Wabnitz, "On the numerical simulation of Kerr frequency combs using coupled mode equations," Opt. Commun. **312**, 134 (2014).

[77] J. P. Gollub, C. W. Meyer, "Symmetry-breaking instabilities on a fluid surface," Physica D: Nonlinear Phenomena **6(3)**, 337-346 (1983).

[78] B. A. Malomed and M. I. Tribelsky, "Bifurcations in distributed kinetic systems with aperiodic instability," Physica D **14**, 67–87 (1984).

[79] P. Coullet, J. Lega, B. Houchmandzadeh, and J. Lajzerowicz, "Breaking chirality in nonequilibrium systems," Phys. Rev. Lett. **65**, 1352 (1990).

[80] M. Kness, L. S. Tuckermann and D. Barkley, "Symmetry-breaking bifurcations in one-dimensional excitable media," Phys. Rev. A **46**, 5054 (1992).

[81] K. Krischer and A. Mikhailov, "Bifurcation to traveling spots in reaction–diffusion systems," Phys. Rev. Lett. **73**, 3165 (1994).

[82] D. Michaelis, U. Peschel, F. Lederer, D. V. Skryabin and W. J. Firth, "Universal criterion and amplitude equation for a nonequilibrium Ising-Bloch transition," Phys. Rev. E **63**, 066602 (2001).

[83] D. V. Skryabin, A. Yulin, D. Michaelis, W. J. Firth, G.-L. Oppo, U. Peschel, and F. Lederer, "Perturbation theory for domain walls in the parametric Ginzburg-Landau equation," Phys. Rev. E **64**, 056618 (2001).

[84] A. W. Liehr, H. U. Bödeker, M. C. Röttger, T. D. Frank, R. Friedrich and H.-G. Purwins, "Drift bifurcation detection for dissipative solitons," New Journal of Physics, **5**, 89 (2003).

[85] S. V. Gurevich, H.U. Bödeker, A. S. Moskalenko, A. W. Liehr, H.-G. Purwins, "Drift bifurcation of dissipative solitons due to a change of shape: experiment and theory," Physica D: Nonlinear Phenomena, **199**, 115-128 (2004).

[86] S. V. Gurevich and R. Friedrich, "Instabilities of Localized Structures in Dissipative Systems with Delayed Feedback," Phys. Rev. Lett. **110**, 014101 (2013).

[87] Y. S. Kivshar and V. V. Afanasjev, "Drift instability of dark solitons in saturable media," Opt. Lett. **21**, 1135-1137 (1996).

[88] C. O. Weiss, H. R. Telle, K. Staliunas, and M. Brambilla, "Restless optical vortex," Phys. Rev. A **47**, R1616(R) (1993).

[89] S. V. Fedorov, A. G. Vladimirov, G. V. Khodova, and N. N. Rosanov, "Effect of frequency detunings and finite relaxation rates on laser localized structures," Phys. Rev. E **61**, 5814 (2000).

[90] G. Izús, M. San Miguel, and M. Santagiustina, "Bloch domain walls in type II optical parametric oscillators," Opt. Lett. **25**, 1454-1456 (2000)

[91] G. J. de Valcárcel, I. Pérez-Arjona, and E. Roldán, "Domain Walls and Ising-Bloch Transitions in Parametrically Driven Systems," Phys. Rev. Lett. **89**, 164101 (2002).

[92] M. Tlidi, A. G. Vladimirov, D. Pieroux, and D. Turaev, "Spontaneous Motion of Cavity Solitons Induced by a Delayed Feedback," Phys. Rev. Lett. **103**, 103904 (2009).

[93] F. Prati, G. Tissoni, L. A. Lugiato, K. M. Aghdami and M. Brambilla, "Spontaneously moving solitons in a cavity soliton laser with circular section," Eur. Phys. J. D **59**, 73–79 (2010).

[94] M. Tlidi, E. Averlant, A. Vladimirov, and K. Panajotov, "Delay feedback induces a spontaneous motion of two-dimensional cavity solitons in driven semiconductor microcavities," Phys. Rev. A **86**, 033822 (2012).

[95] S. B. Ivars, Y. V. Kartashov, Lluis Torner, J. Alberto Conejero, Carles Milián, "Reversible Self-Replication of Spatio-Temporal Kerr Cavity Patterns," Phys. Rev. Lett. **126**, 063903 (2021).

[96] S. B. Ivars, Y. V. Kartashov, P. F. de Córdoba, Lluis Torner, J. Alberto Conejero, Carles Milián, "Photonic snake states in two-dimensional frequency combs," Nat. Photon. **17**, 767–774 (2023).

[97] Q.-T. Cao, H. Wang, C.-H. Dong, H. Jing, R.-S. Liu, X. Chen, L. Ge, Q. Gong, and Y.-F. Xiao, "Experimental demonstration of spontaneous chirality in a nonlinear microresonator," Phys. Rev. Lett. **118**, 033901 (2017).

[98] L. Del Bino, J. M. Silver, S. L. Stebbings, and P. Del'Haye, "Symmetry breaking of counter-propagating light in a nonlinear resonator," Sci. Rep. **7**, 43142 (2017).

[99] M. T. M. Woodley, J. M. Silver, L. Hill, F. Copie, L. Del Bino, S. Zhang, G.-L. Oppo and P. Del'Haye, "Universal symmetry-breaking dynamics for the Kerr interaction of counterpropagating light in dielectric ring resonators," Phys. Rev. A **98 (5)**, 053863 (2018).

[100] I. Hendry, W. Chen, Y. Wang, B. Garbin, J. Javaloyes, G.-L. Oppo, S. Coen, S. G. Murdoch, and M. Erkintalo, "Spontaneous symmetry breaking and trapping of temporal Kerr cavity solitons by pulsed or amplitude-modulated driving fields," Phys. Rev. A **97**, 053834 (2018).

[101] C. Wu, J. Fan, G. Chen, and S. Jia, "Symmetry-breaking-induced dynamics in a nonlinear microresonator," Opt. Express **27**, 28133-28142 (2019).

[102] F. Copie, M. T. M. Woodley, L. Del Bino, J. M. Silver, S. Zhang, and P. Del'Haye, "Interplay of Polarization and Time-Reversal Symmetry Breaking in Synchronously Pumped Ring Resonators," Phys. Rev. Lett. **122**, 013905 (2019).

[103] L. Hill, G. L. Oppo, and P. Del'Haye, "Multi-stage spontaneous symmetry breaking of light in Kerr ring resonators," Commun. Phys. **6**, 208 (2023).

[104] J. Burke and E. Knobloch, "Snakes and ladders: Localized states in the Swift-Hohenberg equation," Phys. Lett. A **360**, 681 (2007).

[105] P. Parra-Rivas, D. Gomila, L. Gelens, and E. Knobloch, "Bifurcation structure of localized states in the Lugiato-Lefever equation with anomalous dispersion," Phys. Rev. E **97**, 042204 (2018).

[106] I. V. Barashenkov and E V Zemlyanaya, "Travelling solitons in the externally driven nonlinear Schrödinger equation," J. Phys. A: Math. Theor. **44**, 465211 (2011).